# Measuring interesting rules in Characteristic rule

Spits Warnars
Department of Computing and Mathematics, Manchester Metropolitan University
John Dalton Building, Chester Street
Manchester M1 5GD, United Kingdom
+44 (0)161 247 1779

s.warnars@mmu.ac.uk

## ABSTRACT
Finding interesting rule in the sixth strategy step about threshold control on generalized relations in attribute oriented induction, there is possibility to select candidate attribute for further generalization and merging of identical tuples until the number of tuples is no greater than the threshold value, as implemented in basic attribute oriented induction algorithm. At this strategy step there is possibility the number of tuples in final generalization result still greater than threshold value. In order to get the final generalization result which only small number of tuples and can be easy to transfer into simple logical formula, the seventh strategy step about rule transformation is evolved where there will be simplification by unioning or grouping the identical attribute. Our approach to measure interesting rule is opposite with heuristic measurement approach by Fudger and Hamilton where the more complex concept hierarchies, more interesting results are likely to be found, but our approach the simpler concept hierarchies, more interesting results are likely to be found and the more complex concept hierarchies, more complex process generalization in concept tree. The decision to find interesting rule is influenced with wide or length and depth or level of concept tree.

## Categories and Subject Descriptors
H 2.8 **[Database Management]** – Database applications – Data Mining; I 2.6 **[Artificial Intelligence]** – Learning – Concept Learning; Induction.

## 1. INTRODUCTION
Attribute oriented induction approach is developed for learning different kinds of knowledge rules such as characteristic rules, discrimination or classification rules, quantitative rules, data evolution regularities [1], qualitative rules [2], association rules and cluster description rules [3]. Attribute oriented induction has concept hierarchy as an advantage where concept hierarchy as a background knowledge which can be provided by knowledge engineers or domain experts [3-5]. Concepts are ordered in a concept hierarchy by levels from specific or low level concepts into general or higher level and generalization is achieved by ascending to the next higher level concepts along the paths of concept hierarchy [8].

DBLearn is a prototype data mining system which developed in Simon Fraser University integrates machine learning methodologies with database technologies and efficiently and effectively extracts characteristic and discriminant rules from relational databases [9,10]. Since 1993 DBLearn have led to a new generation of the system call DBMiner with the following features:

a. Incorporating several data mining techniques like attribute oriented induction, statistical analysis, progressive deepening for mining multiple-level rules and meta-rule guided knowledge mining [11] data cube and OLAP technology [12].
b. Mining new kinds of rules from large databases include multiple level association rules, classification rules, cluster description rules and prediction.
c. Automatic generation of numeric hierarchies and refinement of concept hierarchies.
d. High level SQL-like and graphical data mining interfaces.
e. Client server architecture and performance improvements for larger application.
f. SQL-like data mining query language DMQL and Graphical user interfaces have been enhanced for interactive knowledge mining.
g. Perform roll-up and drill-down at multiple concept levels with multiple dimensional data cubes.

DBMiner had been developed by integrating database, OLAP and data mining technologies[12] which previously called DBLearn have their own database architecture. Concept hierarchy is stored as a relation in the database provides essential background knowledge for data generalization and multiple level data mining. Concept hierarchy can be specified based on the relationship among database attributes or by set groupings and be stored in the form of relations in the same database [11]. Concept hierarchy can be adjusted dynamically based on the distribution of the set of data relevant to the data mining task and hierarchies for numerical attributes can be constructed automatically based on data distribution analysis [11].

For making easy the implementation a concept hierarchy will just only based on non rule based concept hierarchy and just learning for characteristic rule. Characteristic rule is an assertion which characterizes the concepts which satisfied by all of the data stored in database. Provide generalized concepts about a property which can help people recognize the common features of the data in a class. For example the symptom of the specific disease [6].

For doing the generalization there are 8 strategy steps must be done [4], where step 1 until 7 as for characteristic rule and step 1 until 8 for classification/discriminant rule.

a. Generalization on the smallest decomposable components
b. Attribute removal
c. Concept tree Ascension
d. Vote propagation
e. Threshold control on each attribute
f. Threshold control on generalized relations
g. Rule transformation
h. Handling overlapping tuples

## 2. PROBLEM IDENTIFICATION

In the sixth strategy step about threshold control on generalized relations, there is possibility to select candidate attribute for further generalization and merging of identical tuples until the number of tuples is no greater than the threshold value, as implemented in basic attribute oriented induction algorithm [4]. At this strategy step there is possibility the number of tuples in final generalization result still greater than threshold value. In order to get the final generalization result which only small number of tuples and can be easy to transfer into simple logical formula, the seventh strategy step about rule transformation is evolved [4] where there will be simplification by unioning or grouping the identical attribute [2,4]. Based on the above explanation then there are problems like :

a. Which one the best attribute for further generalization ?
b. Which one the best attribute for further simplification ?

Our implementation attribute oriented induction characteristic rule has been implemented with Java programming language and MySQL database with 50.000 records, while data example and concept hierarchy refer to [4,6]. Based on concept hierarchy in [4,6] we have 4 concept trees, they are :

a. Figure 1 is concept tree for major.
b. Figure 2 is concept tree for category.
c. Figure 3 is concept tree for birthplace
d. Figure 4 is concept tree for GPA.

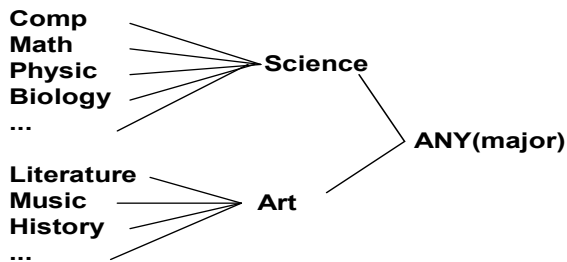

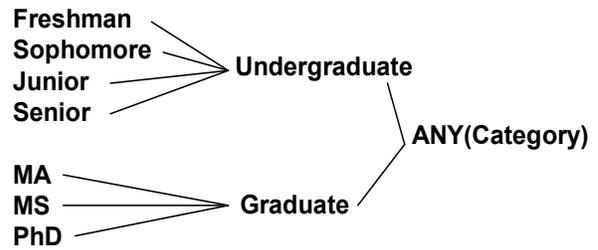

Figure 1. Concept tree for major

Figure 2. Concept tree for category

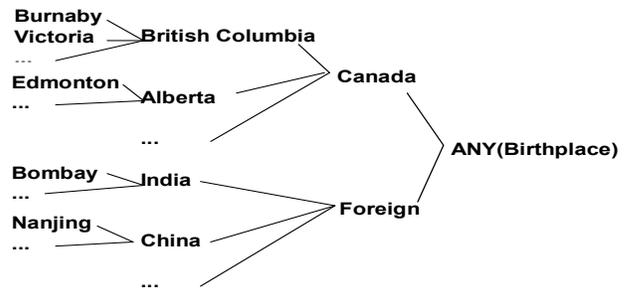

Figure 3. Concept tree for birthplace

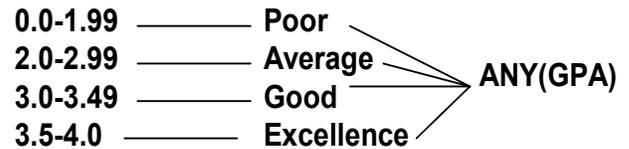

Figure 4. Concept tree for GPA

Figure 5 show the result when program was run to find characteristic rule for graduate student with threshold 2 and stop after the fifth strategy step about threshold control on each attribute.

Figure 5. Result for threshold=2 after threshold control on each attribute

Based on generalization steps for characteristic rule, when the number of distinct tuples still greater than threshold control then the next strategy step which is the sixth strategy step must be done[4]. At explained before because there is possibility the number of tuples still greater then the seventh strategy step must be done. Table 1 until 6 show the possibilities the final generalization include with the rules.

**Table 1. Further generalization on major attribute and unioning on birthplace attribute**

| Major | Birthplace | GPA | Vote |
|---|---|---|---|
| ANY | Canada | {Excellent, Good} | 21599 |
| ANY | Foreign | Good | 10800 |

birthplace(x) ∈ Canada ^ GPA(x) ∈ {Excellent,Good} [66.66%] V

birthplace(x) ∈ Foreign ^ GPA(x) ∈ Good [33.33%]

**Table 2. Further generalization on major attribute and unioning on GPA attribute**

| Major | Birthplace | GPA | Vote |
|---|---|---|---|
| ANY | Canada | Excellent | 17999 |
| ANY | {Foreign,Canada} | Good | 14400 |

birthplace(x) ∈ Canada ^ GPA(x) ∈ Excellent [55.55%] V

GPA(x) ∈ Good [44.44%]

**Table 3. Further generalization on birthplace attribute and unioning on major attribute**

| Major | Birthplace | GPA | Vote |
|---|---|---|---|
| {Art,Science} | ANY | Excellent | 17999 |
| Science | ANY | Good | 14400 |

GPA(x) ∈ Excellent [55.55%] V

major(x) ∈ Science ^ GPA(x) ∈ Good [44.44%]

**Table 4. Further generalization on birthplace attribute and unioning on GPA attribute**

| Major | Birthplace | GPA | Vote |
|---|---|---|---|
| Art | ANY | Excellent | 7200 |
| Science | ANY | {Excellent,Good} | 25199 |

major(x) ∈ Art ^ GPA(x) ∈ Excellent [22.22%] V

major(x) ∈ Science ^ GPA(x) ∈ {Excellent, Good} [77.77%]

**Table 5. Further generalization on GPA attribute and unioning on major attribute**

| Major | Birthplace | GPA | Vote |
|---|---|---|---|
| Art | Canada | ANY | 7200 |
| Science | {Canada, Foreign} | ANY | 25199 |

major(x) ∈ Art ^ birthplace(x) ∈ Canada [22.22%] V

major(x) ∈ Science [77.77%]

**Table 6. Further generalization on GPA attribute and unioning on birthplace attribute**

| Major | Birthplace | GPA | Vote |
|---|---|---|---|
| {Art,Science} | Canada | ANY | 21599 |
| Science | Foreign | ANY | 10800 |

birthplace(x) ∈ Canada [66.66%] V

major(x) ∈ Science ^ birthplace(x) ∈ Foreign [33.33%]

## 3. DEPTH AND LENGTH OF CONCEPT TREE

The final generalization results in table 1 into 6 have the equal interesting rule, the same important and the best result will depend on user's interest. In order to find the best final generalization from six final generalization results in table 1 into 6, where automatically can be built by program application. Our approach to measure interesting rule is influenced by heuristic measurement approach by Fudger and Hamilton where the more complex concept hierarchies, more interesting results are likely to be found [7]. Opposite with Fudger and Hamilton approach, in our approach the interesting rule can be found in the simple concept hierarhies. The simpler concept hierarchies, more interesting results are likely to be found and the more complex concept hierarchies, more complex process generalization in concept tree. The decision to find interesting rule is influenced with wide or length and depth or level of concept tree:

a. Depth or level of concept tree, where simple depth or level in concept tree will have simple generalization process in concept tree, but the more depth or level in concept tree will have more generalization process in concept tree.
b. Wide or length of concept tree or amount of concepts per level in concept tree, where simple concepts will have simple generalization process in concept tree, but the more concepts will have more generalization process in concept tree.

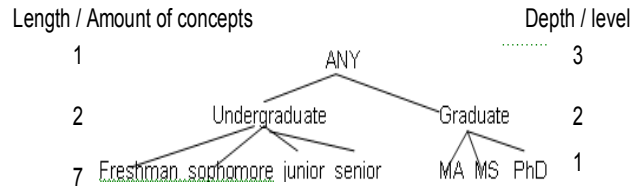

**Figure 6. Depth and length of category concept tree**

For example, figure 6 shows category concept tree has 3 levels and each of level has wide or length of concepts where level 3 as the highest level must have 1 concept, the next level 2 has 2 concepts are undergraduate and graduate and the last level 3 has 7 concepts are Freshman, Sophomore, Junior, Senior, MA, MS, and PhD.

To find the interesting rule based on above explanation then formula (1) will be used to measure concepts in generalization process against concept tree in order to find interesting rule and will be run on each of attribute in process selection generalization process as the sixth strategy step. The simple value as the most interesting value, the highest value as further generalization and the next bigger value as further simplification for unioning or grouping by attribute.

$$(\sum_{i=1}^{n} CR_i/CT_i)/n = (CR_1/CT_1 + ... + CR_n/CT_n)/n \qquad (1)$$

where :

n = Maximum depth /level of concept tree

$CR_i$ = Amount distinct concepts per level in attribute

$CT_i$ = Amount concepts per level in Concept Tree

Table 7 shows the depth and length of concept trees which refer concept hierarchy in [6]

**Table 7. Depth and length of concept tree**

| Depth /level | Length / Amount of Concepts | | | | Total Concepts |
|---|---|---|---|---|---|
| | category | major | birthplace | GPA | |
| 1 | 7 | 11 | 11 | 40 | 69 |
| 2 | 2 | 2 | 5 | 4 | 13 |
| 3 | 1 | 1 | 2 | 1 | 5 |
| 4 | | | 1 | | 1 |
| Total Concepts | 10 | 14 | 19 | 45 | 88 |

For the next explaining will strengthen our approach with variance variable CR as amount concepts in generalization process. The same as before the generalization process for finding characteristic rule for graduate student with threshold value 2 but with different CR as amount of concepts.

Table 8 an example table which shows the result program as shown in figure 5. As a result the highest value as further generalization is birthplace attribute with formula value 0.682 and the unioning based on the next bigger value is major attribute with formula value 0.546 and the interesting attributes is GPA attribute with formula value 0.217 as the lowest value. Thus, the further generalization is on birthplace attribute and unioning on major attribute and table 3 as the interesting generalization relation.

**Table 8. Amount of distinct concepts per level attribute for graduate characteristic with threshold=2**

| | Major | | | birthplace | | | | GPA | | |
|---|---|---|---|---|---|---|---|---|---|---|
| Depth/Level → | 1 | 2 | 3 | 1 | 2 | 3 | 4 | 1 | 2 | 3 |
| CR=Amount concepts | 7 | 2 | | 8 | 5 | 2 | | 6 | 2 | |
| CT=Amount concepts | 11 | 2 | 1 | 11 | 5 | 2 | 1 | 40 | 4 | 1 |
| CR/CT | 0.636 | 1 | 0 | 0.727 | 1 | 1 | 0 | 0.15 | 0.5 | 0 |
| Σ(CR/CT)/n | 1.636/3=0.546 | | | 2.727/4=0.682 | | | | 0.65/3=0.217 | | |

For suppose there is the same highest formula value for attribute major and birthplace as shown in table 9, then there is a problem to decide attribute for further generalization and unioning because of equality formula value. Decision will be based on previous term where simple wide or length and depth or level of concept tree value will have a simple generalization process but in other hand many wide or length and depth or level of concept tree value will have many generalizations processes. As a result because birthplace attribute has 4 levels which more than major attribute with 3 levels then further generalization is on birthplace attribute with formula value 1 and the unioning on the next bigger value is on major attribute with formula value 1 and the interesting attributes is on GPA attribute with formula value 0.75, table 3 for example the result.

**Table 9. Formula execution where there are the same highest formula value**

| | Major | | | birthplace | | | | GPA | | |
|---|---|---|---|---|---|---|---|---|---|---|
| Depth/Level → | 1 | 2 | 3 | 1 | 2 | 3 | 4 | 1 | 2 | 3 |
| CR=Amount concepts | 11 | 2 | 1 | 11 | 5 | 2 | 1 | 10 | 4 | 1 |
| CT=Amount concepts | 11 | 2 | 1 | 11 | 5 | 2 | 1 | 40 | 4 | 1 |
| CR/CT | 1 | 1 | 1 | 1 | 1 | 1 | 1 | 0.25 | 1 | 1 |
| Σ(CR/CT)/n | 3/3=1 | | | 4/4=1 | | | | 2.25/3=0.75 | | |

If suppose the equality value happens on the same level attribute as shown in table 10 where major and GPA attribute have the same level then based on previous term where simple wide or length and depth or level of concept tree value will have a simple generalization process, then the selection will be decided based on multiplication non zero Amount distinct concepts (CR). The highest value multiplication concepts will act as further generalization and the next value as unioning and as result further generalization on GPA attribute where it has result 160 for multiplication 40*4*1, unioning on major attribute where it has result 22 for multiplication 11*2*1 and the interesting attribute is birthplace with formula value 0.859, table 5 for example the result.

**Table 10. Formula execution where there are the same level and highest formula value**

| | Major | | | birthplace | | | | GPA | | |
|---|---|---|---|---|---|---|---|---|---|---|
| Depth/Level → | 1 | 2 | 3 | 1 | 2 | 3 | 4 | 1 | 2 | 3 |
| CR=Amount concepts | 11 | 2 | 1 | 7 | 4 | 2 | 1 | 40 | 4 | 1 |
| CT=Amount concepts | 11 | 2 | 1 | 11 | 5 | 2 | 1 | 40 | 4 | 1 |
| CR/CT | 1 | 1 | 1 | 0.636 | 0.8 | 1 | 1 | 1 | 1 | 1 |
| Σ(CR/CT)/n | 3/3=1 | | | 3.44/4=0.859 | | | | 3/3=1 | | |

If suppose the equality has the same level and multiplication result as shown in table 11 where major and GPA attributes have the same level and multiplication result, then the selection will be decided based on the left or the first attribute. As a result further generalization on major attribute where it has result 22 for multiplication 11*2*1 as the first attribute, unioning on GPA attribute where it has result 22 for multiplication 2*11*1 as the last attribute and the interesting attribute is birthplace with formula value 0.859, table 2 for example the result.

**Table 11. Formula execution where there are the same level and multiplication result**

| Depth/Level → | Major | | | birthplace | | | | GPA | | |
|---|---|---|---|---|---|---|---|---|---|---|
| | 1 | 2 | 3 | 1 | 2 | 3 | 4 | 1 | 2 | 3 |
| CR=Amount concepts | 11 | 2 | 1 | 7 | 4 | 2 | 1 | 2 | 11 | 1 |
| CT=Amount concepts | 11 | 2 | 1 | 11 | 5 | 2 | 1 | 2 | 11 | 1 |
| CR/CT | 1 | 1 | 1 | 0.636 | 0.8 | 1 | 1 | 1 | 1 | 1 |
| Σ(CR/CT)/n | 3/3=1 | | | 3.44/4=0.859 | | | | 3/3=1 | | |

The previous equality value example is on the highest formula value and table 12 is an example when the equality is on the lowest formula value. Based on previous guidance then further generalization on birthplace attribute as highest formula value 0.75 and unioning on GPA attribute with formula value 0.667 which has the highest multiplication amount distinct concepts 160 for multiplication 40*4*1. The interesting attribute is major with formula value 0.667 which has the same value with GPA attribute but has less value multiplication amount distinct concept 22 for multiplication 11*2, table 4 for example the result.

**Table 12. Formula execution where there are the same result at lowest value formula**

| Depth/Level → | Major | | | birthplace | | | | GPA | | |
|---|---|---|---|---|---|---|---|---|---|---|
| | 1 | 2 | 3 | 1 | 2 | 3 | 4 | 1 | 2 | 3 |
| CR=Amount concepts | 11 | 2 | 0 | 11 | 5 | 2 | 0 | 40 | 4 | 0 |
| CT=Amount concepts | 11 | 2 | 1 | 11 | 5 | 2 | 1 | 40 | 4 | 1 |
| CR/CT | 1 | 1 | 0 | 1 | 1 | 1 | 0 | 1 | 1 | 0 |
| Σ(CR/CT)/n | 2/3=0.667 | | | 3/4=0.75 | | | | 2/3=0.667 | | |